\documentstyle[aps,twocolumn,graphicx,ifthen]{revtex}

\newcommand{\myscalebox}[1]{\scalebox{0.4}[0.44]{#1}}
\newcommand{\myscaleboxb}[1]{\scalebox{0.4}[0.4]{#1}}
\newcommand{\mysection}[1]{\section{\em #1}}
\begin{document}
\title{A new method for analyzing ground-state landscapes: ballistic search}

\author{Alexander K. Hartmann\\
{\small  hartmann@theorie.physik.uni-goettingen.de}\\
{\small Institut f\"ur theoretische Physik, Bunsenstr. 9}\\
{\small 37073 G\"ottingen, Germany}\\
{\small Tel. +49-551-395995, Fax. +49-551-399631}}

\date{\today}
\maketitle
\begin{abstract}
A ``ballistic-search'' algorithm
 is presented which allows
the identification of clusters (or funnels) of ground states in Ising spin
glasses even for moderate system sizes. The clusters are defined to be
sets of states, 
which are connected in state-space by chains of zero-energy flips of spins. 
The technique can also be used to estimate the sizes of such clusters.
The performance of the method is tested 
with respect to different system sizes and choices of
parameters. As an application the ground-state funnel
structure of two-dimensional $\pm J$ spin glasses of systems up to size $L=20$
is analyzed by calculating a huge number of ground states per
realization. A $T=0$ entropy per spin of $s_0=0.086(4)k_B$ is obtained.

{\bf Keywords (PACS-codes)}: Spin glasses and other random models (75.10.Nr), 
Numerical simulation studies (75.40.Mg),
General mathematical systems (02.10.Jf). 
\pacs{75.10.Nr, 75.40.Mg, 02.10.Jf}
\end{abstract}

\mysection{Introduction}

The calculation of the energetic minima of spin glass systems
\cite{binder86} remains
the paradigm for difficult 
optimization problems in physics. Usually, only one of
the states exhibiting the lowest energy is calculated, even if a
system is characterized by many minima having all the same lowest energy.
In \cite{alex-valleys} an algorithm is presented, which allows to
analyze large numbers of ground states and enables to
identify all ground-state funnels of Ising spin glass systems
efficiently. Moreover, it is possible to analyze all funnels without
having all ground states available. In this work the algorithm is
presented in detail. Since the algorithm has a random nature, one has
to show that the method is in fact reliable. This will be the main part
of this paper. 
 
The algorithm is applicable to Edwards-Anderson (EA) $\pm J$
spin glasses. They consist of $N$ spins 
$\sigma_i = \pm 1$, described by the Hamiltonian
\begin{equation}
H \equiv - \sum_{\langle i,j\rangle} J_{ij} \sigma_i \sigma_j \quad. 
\end{equation}
The sum $\langle i,j \rangle$ runs over all pairs of neighbors.
The spins are placed on a $d$-dimensional  
lattice of linear size $L$ with periodic boundary conditions in
all directions.
Systems with quenched disorder of the interactions (bonds)
are considered. Their possible values are $J_{ij}=\pm 1$ with equal
probability. To reduce the fluctuations, a  constraint is imposed, so that 
$\sum_{\langle i,j\rangle} J_{ij}=0$.
Since the Hamiltonian exhibits no external
field, reversing all spins of a {\em configuration} (also called {\em
  state}) $z=\{\sigma_i\}$  results in a state with the same energy, 
called the {\em inverse} of
$z$. In the following a spin configuration
 and its inverse are regarded as one single state.

The study of ground-state landscapes helps to understand
the nature of random systems \cite{rieger98}.
But the calculation of the minima of $H$ turns 
out to be a computational hard problem:
It is only for the special case of two-dimensional systems with
periodic boundary conditions in no more than one direction and without
external field that a polynomial-time algorithm is known for obtaining
exact ground states \cite{barahona82b}.  
For more than two dimensions or in the presence of a magnetic field
the problem belongs to the 
class of the NP-hard tasks \cite{barahona82}, 
i.e. only algorithms with exponentially increasing running time
are available. The simplest method works by
enumerating all $2^N$ possible states and has obviously an exponential
running time. Even a system size of $4^3$ is too large.  The basic idea 
of the so
called {\em branch-and-bound} algorithm \cite{hartwig84} is to exclude
those parts of  state space where no low-lying states can be found, so
that the complete low-energy landscape of systems of size $4^3$ 
can be calculated \cite{klotz94}. 

A more
sophisticated method called {\em branch-and-cut} \cite{simone95,simone96}
works by rewriting the quadratic energy function: then  a minimum of
a linear function is to be found, but
an additional set of inequalities  must hold for all feasible solutions.
Since not all inequalities are known a priori, the method iteratively
solves the linear problem, looks for inequalities which are violated,
and adds them to the set until the solution is found. Since the number
of inequalities grows exponentially with the system size the same
holds for the computation time of the algorithm. Anyway, with branch-and-cut
small systems up to $8^3$ are feasible.

The method utilized here is able to
calculate true ground states \cite{alex-stiff} up to sizes $14^3$. 
For two-dimensional systems sizes up to $50^2$ can be
treated. Additionally, in contrast to the
methods mentioned earlier, 
the algorithm used here is able to calculate many statistically independent
ground states for each realization of the randomness.
The method is based on a special genetic
algorithm \cite{pal96,michal92} and on the cluster-exact approximation (CEA)  
method \cite{alex2}. This technique is explained briefly in the next section.

But it is not only the computer time needed for the calculation of one
ground state which may increase exponentially with the system size. 
For the $\pm J$ spin glass 
the number of ground states $D$, called the
{\em ground-state degeneracy}, grows exponentially
with $N$ as well. This is due to the fact that there are
always {\em free} spins, i.e. spins which 
can be turned over without changing the
energy of the system. A state with $f$ independent free 
spins allows for
 $2^f$ different configurations all having the same energy. 
The quantity suitable to describe this behavior is the ground-state entropy
\begin{equation} 
S_0 \equiv k_B \langle \ln D \rangle 
\end{equation}
 where $\langle \ldots \rangle$ denotes 
the average over different realizations of the bonds.
Since the number of free spins is extensive, the entropy per spin
 $s_0\equiv S_0/N$ is non-zero for the $\pm J$ spin glass.

As the ground-state degeneracy increases exponentially, it seems to be 
impossible to obtain all ground states for systems unless they are not very
small. 
To overcome this problem in this work all {\em clusters} 
(also called {\em funnels})
 of ground states are calculated. A cluster is defined in the
 following way: Two ground-state configurations are called {\em
   neighbors} if they differ only by the orientation of one free
 spin. All ground
 states which are accessible through this neighbor relation are
 defined to be in the same cluster.

The method presented here, called {\em ballistic search} (BS),
is able to obtain all
ground-state funnels without knowing all ground states.
Additionally one can  estimate the size of the funnels. 
Consequently, it is possible to calculate directly the ground-state entropy per
spin even for systems exhibiting a huge $T=0$ degeneracy.
Furthermore, the number of funnels and their size-distribution 
as a function of system size are of interest on
their own: for the infinite-ranged Sherrington-Kirkpatrik
(SK) Ising spin glass a complex configuration-space structure was found
using the replica-symmetry-breaking mean-field (MF) scheme by Parisi
\cite{parisi2}. If the MF approximation is valid for finite-dimensional spin
glasses as well, then the number of ground-state funnels must diverge
with increasing system-size. On the other hand the droplet-scaling picture
predicts that, basically, one ground-state funnel dominates
the spin-glass behavior \cite{mcmillan,bray,fisher1,fisher2,bovier}.
To address this issue  a cluster-analysis was performed
for small three-dimensional systems of one size $L=4$ \cite{klotz94}. 
In \cite{garstecki99} two-dimensional spin glasses of sizes up to $5\times 5$
were investigated.
As a first application of BS, an analysis of the size
 dependence of the number of clusters for $d=3$ 
(up to $8\times 8\times  8$) is presented in \cite{alex-valleys}, revealing
 an exponential increase as a function of the number of spins.

The paper is organized as follows: First the procedures
used in this work are presented in detail. 
Then the behavior of the algorithms is tested with
respect to different system sizes and parameters. It is shown that BS
works reliable. In Section IV a
variant of BS is presented, which allows to estimate the size of
clusters, if only a small number of ground states are available per funnel.
Next, as an application, the algorithm is utilized to investigate 
the ground-state cluster structure of
two-dimensional $\pm J$ spin glasses. In particular, the dependence of
the number of clusters and the number of ground states on $N$ are
evaluated. The last section summarizes the results.

\mysection{Algorithms}

First the CEA method is explained briefly. Then, for
illustrating the the problem, a straight-forward method to identify clusters of
ground states is given. In the main part 
the BS method for finding clusters in
systems exhibiting a huge ground-state degeneracy is presented.

The basic method used here for the calculation of spin-glass ground
states is the cluster-exact approximation algorithm 
\cite{alex2}, which is a
discrete optimization method designed especially for spin glasses. In
combination with a genetic algorithm \cite{pal96,michal92} this method is
able to calculate true ground states \cite{alex-stiff} in
three-dimensions for systems of sizes up to $L=14$ on standard workstations. 
A detailed description of the method can be found in \cite{alex-stiff}.
Here the basic ideas of genetic CEA are summarized.

The concept of  {\em frustration} \cite{toulouse77} is important for
its understanding. A system is called frustrated, if it is not
possible to find a configuration, where all bonds contribute with negative
values to the energy. One says it is not possible to {\em satisfy}
all bonds.
The CEA method constructs iteratively and randomly 
a non-frustrated subset of spins within the system.
Spins adjacent to many unsatisfied bonds are more likely to be added to the
subset. During this construction a local gauge-transformation
of the spin variables is applied so that all interactions between subset
spins become ferromagnetic\cite{alex2}.
The  spins not belonging to the subset act like local magnetic fields on the 
subset spins. Therefore,
the ground state of the subset is not trivial.
Since the subset gives raise only to ferromagnetic interactions, 
an energetic minimum state for its spins can be  calculated 
in polynomial time by using graph theoretical methods 
\cite{claibo,knoedel,swamy}: an equivalent network is constructed
\cite{picard1}, the maximum flow is calculated 
\cite{traeff,tarjan} and the spins of the
subset are adjusted to their orientations leading to a minimum in energy
regarding the subset. Therefore, 
the energy is decreased for the total system or remains the same.
By iterating this process a few times the total energy of a system is
decreased quite efficiently, but obtaining ground states turned out to
be very hard.

To increase the efficiency of CEA it is combined with a genetic
algorithm \cite{michal92}.
Genetic algorithms are biologically motivated. An optimal
solution is found by treating many instances of an optimization problem in
parallel, keeping only better instances and replacing bad ones by new
ones (survival of the fittest). With an appropriate choice of few
simulation parameters, usually more than 90\% of all genetic
CEA runs end up with a true ground state. Configurations with a higher
energy are not included in further calculations.

Using this method one does not encounter ergodicity problems or critical
slowing down like in algorithms which are based on Monte-Carlo methods. 
Moreover, it is possible to calculate many statistically
independent configurations (replicas). Genetic CEA
was already utilized to examine the ground-state landscape of
two-dimensional \cite{alex-2d} and three-dimensional 
\cite{alex-sg2} $\pm J$ spin glasses by calculating a small
number of 
ground states per realization. Furthermore the existence of a 
spin-glass phase for nonzero temperature was confirmed for the
three-dimensional spin glass \cite{alex-stiff}. Finally, the method
 was applied to the  $\pm J$ random-bond model to investigate its
$T=0$ transition from ferromagnetism to spin-glass behavior 
\cite{alex-threshold}, which takes place by increasing the fraction of
antiferromagnetic bonds starting from a ferromagnet.

Once many ground states are calculated the straight-forward method to
obtain the structure of the clusters works as follows: 
the construction starts
with one arbitrarily chosen ground state. 
All other states, which differ from this state by one free spin,
are said to be its neighbors. They are added to the
cluster. These neighbors are treated recursively in the same way: all
their neighbors which are yet not included in the cluster are
added. After the construction of one cluster is complete, the
construction of the next one starts with a ground state, which has not
been visited so far.

The running-time of the 
construction of the clusters is only a linear function
of the degeneracy $D$ ($O(D)$), similar to the Hoshen-Kopelman
technique \cite{hoshen76}, because each ground state is visited only
once. Unfortunately, the detection of all neighbors, which has to be 
performed at the beginning, is of $O(D^2)$,
because all pairs of states have to be compared. Since for $L=5$
systems it is possible that they exhibit more than $10^5$ ground states, this
algorithm is not suitable for larger sizes than $L=5$

The ballistic-search method instead is able to treat larger systems.
Its basic idea  is to use a {\em test}, which
tells whether two ground states are in the same cluster or not. The test works
as follows: Given two independent replicas $\{\sigma_i^{\alpha}\}$
and $\{\sigma_i^{\beta}\}$ let $\Delta$ be the set of spins, which are
different in both states: $\Delta\equiv \{i|\sigma_i^{\alpha}\neq
\sigma_i^{\beta}\}$. Now BS tries to build a path in configuration-space
of successive
flips of free spins, which leads from  $\{\sigma_i^{\alpha}\}$ to
$\{\sigma_i^{\beta}\}$. The path consists of states which differ only
by flips of free spins from $\Delta$ (see Fig. \ref{figAlgo}).
For the simplest version iteratively a free spin is selected randomly
from $\Delta$, flipped
and removed from $\Delta$. This test does not guaranty to find a path
between two ground states which belong to the same cluster. It
may depend on the order of selection of the spins whether a path is
found or not, because not all free spins are independent of each other.
Thus, a path is found with a certain probability $p_f$, 
which depends on the size of $\Delta$. Later on the behavior of
$p_f$ is analyzed.

The algorithm for the identification of clusters using BS works as
follows: the basic idea is to let a ground state represent that part
of a funnel which can be reached using BS with a high probability 
by starting at this ground state. If a
cluster is large it has to be represented by a collection of states, 
so that the whole
cluster is ``covered''. For example a typical cluster of a $L=8$ spin
glass consisting of $10^{17}$ ground states is usually represented by
only some few ground states (e.g. two or three). 
A detailed analysis of how many representing
ground states are needed as a function of cluster and system size
can be found in the next section. At each time the
algorithm stores a set of $m$ clusters
$A=\{A(r)|r=1,\ldots,m\}$ 
each consisting of a set $A(r)=\{z^{rl}\}$ of
representing configurations 
$z^{rl}=\{\sigma^{rl}_{i}\}$ $(l=1,\ldots,|A(r)|)$. 
At the beginning the cluster-set is empty.
Iteratively all available ground states $z^j=\{\sigma^j_i\}$
($j=1,\ldots,D$) are treated: The BS algorithm tries to find paths from $z^j$
or its inverse to all representing configurations in $A$.
Let $F$ be the set of clusters-numbers, where a path is found. Now three
cases are possible (see Fig. \ref{figThreeCases}):
\begin{itemize}
\item No path is found: $F=\emptyset$\\
A new cluster is
created, which is represented by the actual configuration treated:
$A(m+1)\equiv \{z^j\}$. The cluster is added to $A$: $A\equiv A\cup
\{A(m+1)\}$.
\item One or more paths are found to exactly one cluster: $F=\{f_1\}$.
  Thus, the ground state $z^j$ belongs to one
  cluster. Consequently, nothing special happens, the set $A$ remains
  unchanged.
\item $z^j$ is found to be in more than one cluster: 
$F=\{f_1,\ldots,f_k\}$. All
these clusters are merged into one single cluster, which is now represented by
the union $\tilde{A}$
of the states, which have represented before all clusters affected by
the merge:\\
$\tilde{A}\equiv \bigcup_{j=1}^k A(f_j)$, 
$A\equiv \{\tilde{A}\}\cup A \setminus \bigcup_{j=1}^k \{A(f_j) \}$
\end{itemize} 

The whole loop is performed two times. The reason is that a state
which links two parts of a large cluster (case 3) may appear in the
sequence of ground states before
states appear belonging to the second part of the cluster. 
Consequently, this linking
state is treated as being part of just one single smaller cluster and both
subclusters are not recognized as one larger cluster 
(see Fig. \ref{figTwoRuns}). During the second iteration the
``linking'' state is compared to all other representing states found
in the first iteration, i.e. the large cluster is identified
correctly. With one iteration, the problem
appears only, if few ground-states per cluster are
available. Nevertheless, always two iterations are performed, so the
difficulty does not occur.

The BS-identification algorithm has the following advantages:
since each ground-state configuration represents many ground
states, the method does not need to compare all pairs
of states. Each state is compared only with the representing
configurations. For the system sizes usually encountered, this value
is only slightly larger than the number of funnels itself.
Thus, the computer time needed for the calculation 
grows only a little
bit faster than $O(Dn_C)$, where $n_C$ is the number of clusters.
 Consequently, large sets of ground states, 
which appear already for small system sizes, can be treated. Furthermore, the
ground-state funnel-structure of even larger
systems can be analyzed, since it is sufficient that there are only a small
number of ground states per cluster available. One has to ensure that
really all clusters are found, which is simply done by calculating 
enough states. A study of how many states are needed for different
sizes $L$ is presented in the next section.

\mysection{Numerical Tests}

Since the BS cluster algorithm does not guaranty to find all clusters,
numerical tests on two- and three-dimensional systems were performed.
Here, the tests for three dimensions are presented, because for this
type of
system results are already available \cite{alex-valleys}. For two
dimensions the algorithm behaves similar. Results concerning the
number of ground states and the number of funnels for $d=2$ are
presented later on.

For system sizes $L=3,4,5,6,8$ large numbers of independent ground
states were calculated using genetic CEA. Usually 1000
different realizations of the disorder were considered. Tab. I shows
the number of realizations $n_R$ and the
number of independent runs $r$ per realization for 
different system sizes $L$. For small systems sizes (and for 100
realizations of $L=5$) many runs plus an additional local search
were performed to calculate all
existing ground states. For the larger sizes $L=5,6,8$ the number of
ground states is too large, so it is only possible to try to
calculate at least one ground state per cluster. We will see later
that for most of the realizations it is highly probable that all
existing clusters were found using genetic CEA.

But first we concentrate on another issue: The
ground states were grouped into clusters using the ballistic-search
algorithm. To interpret the following results correctly, one
should keep in mind that for detecting  one ground state
being part of a cluster it is sufficient to find just one path to any
of the other states of the cluster. 
The question under consideration now is:
how large is the probability, that for ground states belonging
to the same cluster  the BS test finds a  path.

To investigate this question the following test was performed: Many
thousand times pairs of ground states were selected, which belong to
the same cluster. The probability for selecting a pair was proportional
to the size of the cluster (How to estimate the size of a cluster,
if not all ground states are available, is shown in the next section).
This guarantees that each ground state contributes to the result with
its proper thermodynamical weight.
The outcome of this test depends on the assumption, that the 
construction of the
clusters has been performed correctly. Later we will see, that 
this  has indeed been the case with a very high probability.
Let $p_f$ be the probability that BS finds a direct path in
configuration space connecting two given states. The result is expected to
depend on the number of spins, which are different in both states,
i.e. on the length $l_{\mbox{\small path}}$ of the path. The result is shown
in Fig. \ref{figPWayLength} for system sizes $L=3,4,5,6,8$. 
 The probability decreases with increasing length of the path. Thus, finding
a successful path becomes more difficult, which is to be expected,
since the number of possible paths 
increases exponentially with $l_{\mbox{\small path}}$. On the other hand, by
increasing the system size, it is more likely to find
 a connecting path.
This is caused by the fact that the number of isolated free spins
increases, which in fact can be flipped in any order. To investigate the
dependence on $L$ the following finite-size behavior is assumed
($\lambda$ being a scaling exponent):

\begin{equation}
p_f(l_{\mbox{\small path}}, L) = 
\tilde{p}_f(L^{-\lambda}l_{\mbox{\small path}}) \quad .
\end{equation}

By plotting $p_f(l_{\mbox{\small path}}, L)$ against
$L^{-\lambda}l_{\mbox{\small path}}$
with correct parameter $\lambda$ the datapoints for different system
sizes near $l_{\mbox{\small path}}=0$
should collapse onto a single curve. The best results were obtained
for $\lambda=1.7$. In Fig. \ref{figPWayScale} the resulting scaling plot
is shown. Now assume that two ground states differ by a certain
fraction of spins. Thus, the absolute number of spins being different
increases with $L^3$. Since the length of a path for a fixed value of
$p_f$ increases only with $L^{1.7}$, it becomes indeed more and more difficult
to find a path with increasing system size $L$.

So far the behavior of the ballistic search was investigated. But
what does it mean for the cluster-identification algorithm? We are
interested in the question, whether all clusters are identified
correctly.  This can be formulated as a generalized percolation problem:

\begin{itemize}
\item Consider
\begin{itemize}
\item A set $B=\{z^1,\ldots,z^K\}$ of objects
\item A distance function $d(z^a,z^b)$
\item A probability $p_{\mbox{\small bond}}(d)$, that a bond is created between
  two elements from
  $B$, The probability depends on the distance $d$ between the elements and
 decreases monotonically with $d$.
\end{itemize}
\item The quantity of interest is the probability $p_1$ that 
  {\em all} objects belong to the same
  cluster, i.e. the probability that there is only one cluster.
\end{itemize}

One can identify $B$ with a set of ground states belonging all to the same
 cluster, $d(z^a,z^b)$ with $l_{\mbox{\small path}}$ and 
$p_{\mbox{\small bond}}(d)$ with
the probability $p_f$ that a path is found. Then the quantity $p_1$ is the
probability that all ground states are identified correctly 
by the BS clustering algorithm as being members of the same cluster.
Since the average distance between different states decreases for a
given cluster by increasing the number $K$ of states, 
$p_1$ should increase with
$K$. The reason is that more bonds are likely to be created 
(see Fig. \ref{figMoreStates}). 
It should be possible to determine
$p_1(K)$ for different functions $d(z^a,z^b)$ and
$p_{\mbox{\small bond}}(d)$, at least numerically. But here a 
different approach
is selected: Since all ground states and funnels are available, $p_1(K)$
can be evaluated directly. For each realization, each lattice size $L$ and
each number $k\in[2,20]$, a set of $K$ different ground states
was selected 50 times randomly from one cluster. 
Again each ground-state
funnel was chosen with a weight proportional to its size. The BS
clustering method was applied and it was verified whether just one
cluster was found. The result is shown if Fig. \ref{figPOneNum}. As the
system size increases, larger clusters occur, which  are
harder to identify. But is
visible that even two ground states are sufficient most of the time to
identify a cluster correctly. To be almost sure, 
a value of $p_1>0.999$ is expected to be
sufficient, which means that $K=10$ is enough for $L=3$ and, as found
by further analysis, $K=40$ for $L=8$.
 
In fact, for the largest ground-state funnels found in this work
there are usually many more ground
states available than needed for identifying a cluster correctly
with a probability of 99.9\%. Consequently, our results for the
probability $p_f$ and $p_1$ are very reliable.
Furthermore, whenever the number of ground states is too
small, it is always possible to generate additional states by
performing $T=0$ Monte-Carlo simulations, i.e. selecting
spins randomly and flipping them if they are free. 
Consequently, we can be sure that the
funnels, which were used for the preceding analysis, were obtained
correctly. 

Another question is, whether for a given realization there are some
ground-state funnels, for which no ground states are found using
just a restricted number of runs of genetic CEA. 
This problem does not occur for the smallest sizes,
because it is possible to calculate all states of lowest energy using
that method. But even for $L=6$ there are realizations already exhibiting more
than $10^6$ ground states, making it impossible to obtain all of them
directly. The genetic CEA method calculates a ground state with a
probability $p_C$,  which increases on average with the size $|C|$ 
of the cluster 
it belongs to \cite{alex-eig-cea}. Thus, ground states belonging to
small funnels have a small probability $p_C$ of being found using a
finite number of runs. This probability is not extremely small, since
$p_C$ increases slower than the size of a cluster 
$|C|$ \cite{alex-eig-cea}, i.e. for $|C_1|<|C_2|$

\begin{equation}
\frac{p_{C_1}}{|C_1|} > \frac{p_{C_2}}{|C_2|} \label{eq_prob_raise}
\quad ,
\end{equation} 
but it is still small enough that some funnels may have been missed.
Since $p_C$ increases with $|C|$, the probability that a cluster is
{\em not} found at all takes the largest values for the
smallest clusters, i.e. for a cluster of size 1. This probability is
denoted here with  $\overline{p}_1$. Consequently, $\overline{p}_1$ 
is an upper limit for the probability that a certain cluster is
missed. Now $\overline{p}_1$ is estimated.

Consider a list of $K$ ground states $z^1,\ldots,z^K$ ($z^j=\{\sigma^j_i\}$),
in the order they were obtained in the calculation using genetic
CEA. Thus, on average states from larger funnels appear earlier in the
list than states from smaller funnels. A state which is calculated
several times, is stored in the list just once. For each state
the number of times $h_j$ it has occurred is recorded, 
let $h\equiv\sum_j h_j$.  For small systems, where the
number of existing ground states is small compared to the number of
runs, usually $h_j>1$. Now we look at the smallest cluster $C_{\min}$,
which was found using the procedure. The relative frequency, that
a ground state from $C_{\min}$ is found, is approximately
$p_{\min}=\sum_{j\in C_{\min}}h_j/h$. It follows from
(\ref{eq_prob_raise}) that $p_1 > p_{\min}/|{C_{\min}}|$. Thus, we
have for the probability $\overline{p}_1$
that a cluster of size 1 is not found during  $h$
different runs 
\begin{equation}
\overline{p}_1 =(1-p_1)^h <
\left(1-\frac{p_{\min}}{n_{C_{\min}}}\right)^h \label{eq_p1} \quad .
\end{equation}

Consequently, it is possible to estimate for each single configuration
the likelihood that a small cluster may have been missed. 
For the smaller sizes,
where it was claimed that all ground states were found using a large
number of runs, typical values $\overline{p}_1<10^{-10}$ were found. A
small cluster was missed with $\overline{p}_1>0.01$ only
for three realizations of $L=3$ and never for $L=4,5$. Thus, it is
highly probable, that all ground states were found for the smaller
sizes $L=3,4,5$

For larger sizes, the number of states obtained per cluster is small
compared to the size of the cluster. The estimate (\ref{eq_p1})
gives always a large value. Consequently, another method of estimating
the quality of the results has to be applied: the progression of the
BS cluster algorithm is observed during the processing of the ground
states $z^1,\ldots,z^K$. Each state which causes 
a new cluster to be created or some
clusters to be  merged is called an {\em event}. Since there is only a
finite number $n_C$ of clusters and each cluster is represented by a finite
number of configurations there is only a finite number of
events. For the systems sizes encountered here, the number of events
is only slightly larger than $n_C$, because most of the clusters are
represented by just one ground state. In principle, if the last event is
known, no further ground states have to be processed. Since the last
event is not known for system sizes $L>5$, one can only assume 
that the last event has already occurred, 
if no new event is found for a long time, while treating more and more
states $z^j$. At each step let

\begin{equation}
Q(j)\equiv \frac{j}{\mbox{number of the last event before}\, z^j}
\quad .
\end{equation}

The fraction Q measures the relative length of sequences, where no event
occurs. For $j$ beyond the last event, $Q(j)\to\infty$.
The longest sequence found before the last event
$Q\equiv \max_{j\le \mbox{\small last}} Q(j)$,
describes how many states are needed to find all clusters
without knowing the last event. 

Using the number of runs of the genetic CEA algorithms given above,
a value of Q larger than 4 was never observed for $L=3$. This means, one
can be sure that all funnels have been identified, if the number of
ground states processed is four times larger than the number of the
state constituting the last event. For $L=4$ the
largest $Q$ found was 3 and for $L=5$ it was observed that 
$Q<2.5$ for all realizations. 
For larger sizes Q is not known, so $Q(K)$ is used instead. Assume
that the
last event has already occurred, then by
including more and more states into the analysis, $Q(K)$ grows
linearly and the confidence increases that really no further event is to be
expected.
We believe that if the number $K$ of ground states is more than four
times the number of the last event in $z^1,\ldots,z^K$, i.e. $Q(K)>4$ 
one can be quite sure that all
funnels have been found, because $Q\le 4$ for all small sizes.
 Even if $Q(K)=2$, it seems very likely that
no cluster has been missed, since $Q$ seems to decrease by going to
larger sizes. For $L=5,6$  $Q(K)<4$ was found for only about 8\% of the
realizations, and $Q(K)<2$ only for 2\% ($L=5$) respectively
5\% ($L=6$). Consequently, nearly
all clusters have been identified. 
$Q>4$ holds for 75 \%
of all $L=8$ realizations  ($Q>2$ for 85\%), i.e. here a small number
of funnels may have
been missed for 25\% of all realizations, while for the
majority of the realizations really all funnels have been detected.
Only by increasing massively the number of available 
ground states per realization, a substantial improvement 
for the largest size treated in this work is possible.

On the other hand, if physical properties have to be evaluated, the
results are very reliable even for $Q(K)\approx2$, 
because each cluster contributes
with a weight proportional to its size. As mentioned before, the
probability that genetic CEA returns a certain ground state increases
with cluster size. Consequently, only small clusters are omitted and the
result is affected only slightly.

\mysection{Size of a cluster}

Once all clusters are identified their sizes have to be obtained to calculate
the entropy. A variant of BS is used to perform this task. Starting from a
state $\{\sigma_i\}$ from a cluster $C$,
free spins are flipped iteratively, but each spin not more than
once. During the iteration additional free spins may be generated and
other spins may become fixed.
When there are no more free spins left the process stops.
Thus, one has constructed a straight path in state space from the
ground state
to the border of the funnel $C$. The number of spins that has been flipped 
is denoted by $l_{\max}$. By averaging
over several trials and several ground states of a cluster, one obtains an
average value $\overline{l}_{\max}$, which is a measure for the size
of the cluster.

For system sizes $L=3,4,5$ all ground states were available (for
100 realizations for $L=5$) and the cluster sizes are known exactly.
Fig. \ref{figEstSize} displays the average size $V$ of a cluster as a
function of $l_{\max}$. An exponential dependence is
found, yielding $V=2^{\alpha l_{\max}}$ with $\alpha=0.90(5)$. The
deviation from the pure exponential behavior for the largest clusters
of each system size should be a finite size effect. 

One might think that instead of successively turning spins over, one
could simply count the static number of free spins. But it turns out
that the quantity $l_{\max}$ describes the size of a cluster better.
The reason is that by flipping spins
additional free spins are created and deleted. Consider for example a
one-dimensional chain of $N$ 
ferromagnetic coupled spins with antiperiodic boundary
conditions. Each ground state consists of two linear domains of spins. In
each domain all spins have the same orientation. For each
ground-state there are just two free spin, but
all $2N$ ground states belong to the same cluster. The possibility of
similar ground-state topologies  is 
taken into account using the definition given above.

\mysection{Results}

The data presented in the preceding sections show that earlier
results \cite{alex-valleys} for three-dimensional spin glasses are
reliable, where an exponential increase of the degeneracy and the number
of ground-state funnels was found. In this section a similar analysis
of the ground-state landscape of two-dimensional systems is performed.
It will be shown that qualitatively the behavior is the same as for $d=3$.

For system sizes $L=5,7,10,14,20$ large numbers of independent ground
states were calculated using genetic CEA, up to $10^4$ runs per
realization were performed. Since many runs are needed
to describe the ground-state landscape as completely as possible, no
runs for larger systems were conducted, although it is possible to
obtain true ground states easily up to $L=50$. Usually 1000
different realizations of the disorder were considered, except for $L=20$,
where only 92 realizations could be treated.  
For the small systems sizes $L=5,7 $ many runs plus an additional local search
were performed to calculate {\em all}
ground states. For the larger sizes $L=10,14,20$ the number of
ground states is too large, so we restrict ourselves to
calculate at least one ground state per cluster. The probability that
some clusters were missed is higher for two dimensions than for the
$d=3$ case, because the ground-state degeneracy grows faster with the
system size: for small systems sizes $L\le 10$
it is again highly probable that all funnels
have been obtained. For $L=14$ some small
funnels may have been missed for about 30\% of all realizations, 
while for $L=20$ this fraction raises even to
60\%. This is due to the enormous computational effort needed for the 
largest systems. 
For the $L=20$ realizations a total computing time of more than 2
CPU-years was consumed on a cluster of Power-PC processors running with
80MHz.

The
ground states were grouped into clusters using the ballistic-search
algorithm. The number of states per funnel was sufficiently large, so
that only with a probability of less than $10^{-3}$ some configurations
from a large cluster may be
mistaken for belonging to  different funnels.
In Fig. \ref{figNumClusters} the average number $n_C$ of clusters is
shown as a function of the number  $N$ of spins. 
By visualizing the results in a double-logarithmic plot (see
inset) one realizes
that $n_C$ seems to grow faster than any power of $N$. The larger
slope in the linear-logarithmic plot 
for small systems may be a finite-size effect. Additionally for
$L=20$ there is a large probability that some small funnels are missed,
explaining the smaller slope there.
Consequently, the data presented here favor an 
exponential increase of $n_C(N)$.

For the small systems sizes the number of ground states
in each cluster could be counted directly. For the larger sizes the
variant of the BS method was used to estimate the size of each
cluster. The  average size $V$ of a cluster as a
function of $l_{\max}$ is displayed in Fig. \ref{figEstSizeSquare}.
Similar to the results presented in the preceding section, 
an exponential dependence is
found, yielding $V=2^{\alpha l_{\max}}$ with $\alpha=0.85(5)$.

By summing up all cluster
sizes for each realization the ground-state degeneracy $D$ is obtained.
The quantity averaged over all realizations 
is shown in Fig. \ref{figNumStates}
as a function of $N$. The exponential growth is obvious. 

The result for the average ground-state entropy per spin is 
shown in the inset of Fig.  \ref{figNumStates}. By fitting a function of the
form $s_0(L)=s_0(\infty)+a*L^{-\beta}$ a value of 
$s_0(\infty)=0.0856(4)k_B$ is obtained. 

In \cite{morgenstern80b} 
$s_0\approx 0.075k_B$ was estimated by using a recursive method to
obtain numerically exact free energies up to $L=18$. 
For technical reasons, only systems with periodic
boundary conditions in one direction were treated, 
which may be the reason for the smaller result. The result of
$s_0\approx 0.07k_B$ found in \cite{vannimenus77} is even slightly lower.
The value found by a Monte-Carlo
simulation $s_0\approx0.1k_B$ \cite{kirkpatrick77} for systems of size
$80^2$ is much larger. The deviation is presumably caused by the fact
that it was not possible to obtain true ground states for systems of
that size, i.e. too many states were visited.
The entropy is considerable higher than for the three-dimensional $\pm
J$ spin glass, where $s_0(\infty)=0.051(1)k_B$ was obtained using the
same method applied here \cite{alex-valleys}.

The result for the entropy does not suffer from the fact, that some
ground-state funnels may have been missed for $L=14,20$: the
probability for finding a cluster 
by applying genetic CEA grows with the size of
the cluster \cite{alex-eig-cea}. 
This implies that the clusters which may have been missed are
considerably small, so the influence on the result is
negligible. The largest source of uncertainty is caused by the
assumption that the size of a cluster grows like $2^{\alpha
  \overline{l}_{\max}}$. The error of the constant $\alpha$ enters linearly 
into the result of the entropy. 
To estimate the influence of this approximation, $s_0$ was
calculated using estimated cluster sizes as well for the three smallest
systems sizes, where the entropy had been obtained exactly. For both
cases the results were equal to the exact values within error bars.
The final result quoted here is $s_0=0.086(4)$.

\mysection{Conclusion}

The ballistic-search method has been 
presented, which
allows the fast identification of very large clusters, appearing for
example in the calculation of the ground-state landscape of $\pm J$
spin glasses. Furthermore, it is possible
to calculate clusters of systems when only a small fraction of 
all their states is available. The method should be extendable to similar
clustering problems, especially for analyzing results from simulations
at finite temperature. A variant of the technique is used to estimate the
size of the clusters. 

Since the BS algorithm does not guarantee to find a path in configuration space
between two ground states which belong to the same cluster, extensive
numerical tests were performed. It was shown that by increasing the
number of available states, it is possible to reduce the probability
that a cluster is not identified correctly. Additionally
it is possible to estimate the probability that small clusters are not
found. Consequently, the new technique enables to 
analyze the complete funnel structure for 
two-dimensional $\pm J$ spin glasses up to $L=20$ and for
three-dimensional systems
 up to $L=8$. Thus, systems exhibiting up to $10^{17}$
ground states can be treated efficiently.

For $d=2$ an analysis of the ground-state landscape has been presented.
The number of funnels and the ground state degeneracy
increases exponentially with the system size.  
The ground-state
entropy per spin was found to be $s_0=0.086(4)k_B$. Results for three
dimensional systems can be found in \cite{alex-valleys}. It should be
pointed out that the result for the entropy does not depend on the
way a cluster is defined. The specific definition given here is only a
tool which allows the treatment of systems exhibiting a huge
ground-state degeneracy.

\mysection{Acknowledgements}

The author thanks K. Battacharya for interesting discussions. He is
grateful to P. M\"uller for critically reading the manuscript.
He was supported by the Graduiertenkolleg
``Modellierung und Wissenschaftliches Rechnen in 
Mathematik und Naturwissenschaften'' at the
{\em In\-ter\-diszi\-pli\-n\"a\-res Zentrum f\"ur Wissenschaftliches
  Rechnen} in Heidelberg and the
{\em Paderborn Center for Parallel Computing}
 by the allocation of computer time.  The author obtained financial
 support from the DFG ({\em Deutsche Forschungs Gemeinschaft}).

\newcommand{\captionAlgo}
{Ballistic search: A path 
  in configuration-space of free spins is constructed between two
  ground states (dark nodes) belonging to the same
 cluster. Depending on the order the spins are
  flipped the path may be found or not. Nodes represent ground
  states, edges represent flips of free spins. Please note that this figure is
  only a cartoon, the configuration space has $N$ dimensions.}

\newcommand{\captionThreeCases}
{Algorithm for the identification of all clusters: several
ground-states (circles) ``cover'' parts of clusters (filled areas). During
the processing of all states a set of clusters is kept. When state
$z^j$ is treated, it is tested using BS to how many of the already
existing clusters the state belongs. Three cases can occur: 
a) The ground state is found
to belong to no cluster, b) it is found in exactly
one cluster and c)
it is found in several clusters. In the first case a new cluster is
found, in the second one nothing changes and in the third case several
smaller clusters are identified as subsets of the same larger cluster.}

\newcommand{\captionTwoRuns}
{Example, that the order the states are treated may affect the
  result. Consider three states $z^1,z^2,z^3$, all belonging to the same
  cluster. Assume that BS finds a path between $(z^1,z^2)$ and
  $(z^2,z^3)$, but not between $(z^1,z^3)$.
In the first case two clusters are found (false), in the second case
 one cluster (correct). To obtain always the correct result,
 two iterations are  needed.
}

\newcommand{\captionPWayLength}
{Probability $p_f$ that
BS finds a path of free flipping spins between two ground
  states belonging to the same funnel of three-dimensional $\pm J$
  spin  glasses. $p_f$ is shown as a function
  of the number $l_{\mbox{\small path}}$ of spins, 
which are different in the two states   for lattice sizes
$L=3,4,5,6,8$.}

\newcommand{\captionPWayScale}
{Finite-size scaling plot of $p_f(L,l_{\mbox{\small path}})$ (see
  Fig. \ref{figPWayLength}). A scaling behavior
of $p_f(L,l_{\mbox{\small path}}) = \tilde{p}(L^{-\lambda}l_{\mbox{\small path}})$ is
assumed. Using $\lambda=1.7$ the data points for $L=4,5,6,8$ fall onto
one curve near the origin $l_{\mbox{\small path}}=0$.}

\newcommand{\captionMoreStates}
{By increasing the number of available states per cluster, the
  probability  increases that they are identified as members of the same 
  cluster. Circles denote states, the thickness of the line
  represents the probability that a path is found by the ballistic search.}

\newcommand{\captionPOneNum}
{Probability $p_1$ that a sample of $K$ ground states belonging to the
  same cluster is indeed identified by the BS
  clustering algorithm as one single cluster as a function of $K$ for
  different system sizes $L$ ($d=3$).}

\newcommand{\captionEstSize}
{Average size $V$ of a cluster as a function of average dynamic number
  $l_{\max}$ of
  free spins (see text) for three-dimensional $\pm J$ spin glasses of
 system sizes $L=3,4,5$, where all ground have been obtained. A
 $V=2^{0.9l_{\max}}$ dependence is found, indicated by a line.}

\newcommand{\captionNumClusters}
{Number $n_C$ of ground-state clusters for two-dimensional $\pm J$ spin
  glasses as a function of system size
  $N$. The inset shows the same data using 
  a double-logarithmic scale. Lines are guide to the eyes only.}

\newcommand{\captionEstSizeSquare}
{Average size $V$ of a cluster as a function of average dynamic number
  $l_{\max}$ of
  free spins (see text) for two-dimensional $\pm J$ spin glasses of
 system sizes $L=5,7$, where all ground have been obtained. A
 $V=2^{0.85l_{\max}}$ dependence is found, indicated by a line.}

\newcommand{\captionNumStates}
{Number $D$ of ground-states for two-dimensional $\pm J$ spin
  glasses as a function of system size $N$. The
  number of states grows exponentially with the number of spins. Lines
  are guide to the  eyes only. The inset displays the ground-state
  entropy per spin as a function of $L$. The line shows a fit
  extrapolating $s_0$ to the infinite system which yields
$s_0(\infty)=0.0856(4)k_B$.}

\begin{figure}[htb]
\begin{center}
\myscaleboxb{\includegraphics{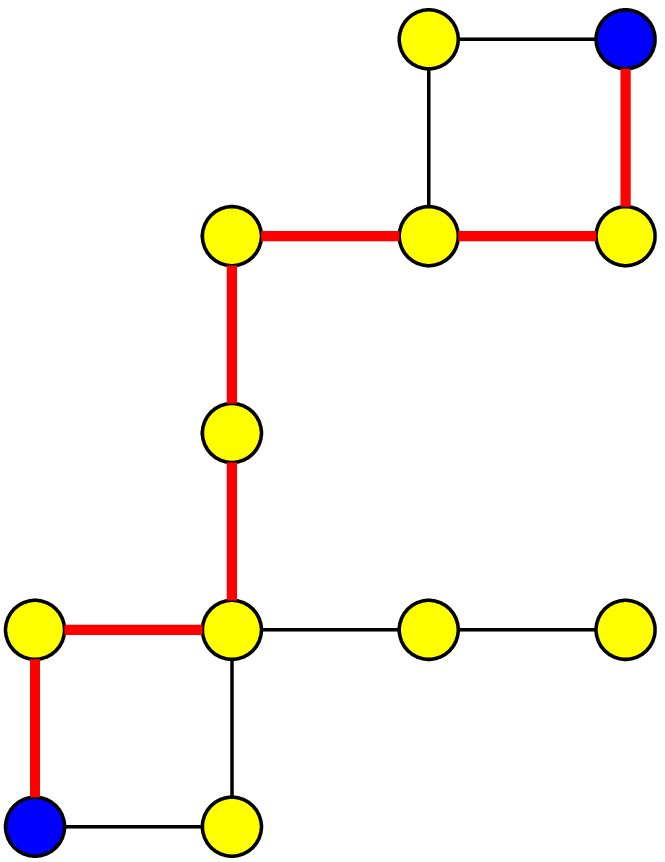}}
\end{center}
\caption{\captionAlgo}
\label{figAlgo}
\end{figure}

\begin{figure}[htb]
\begin{center}
\myscaleboxb{\includegraphics{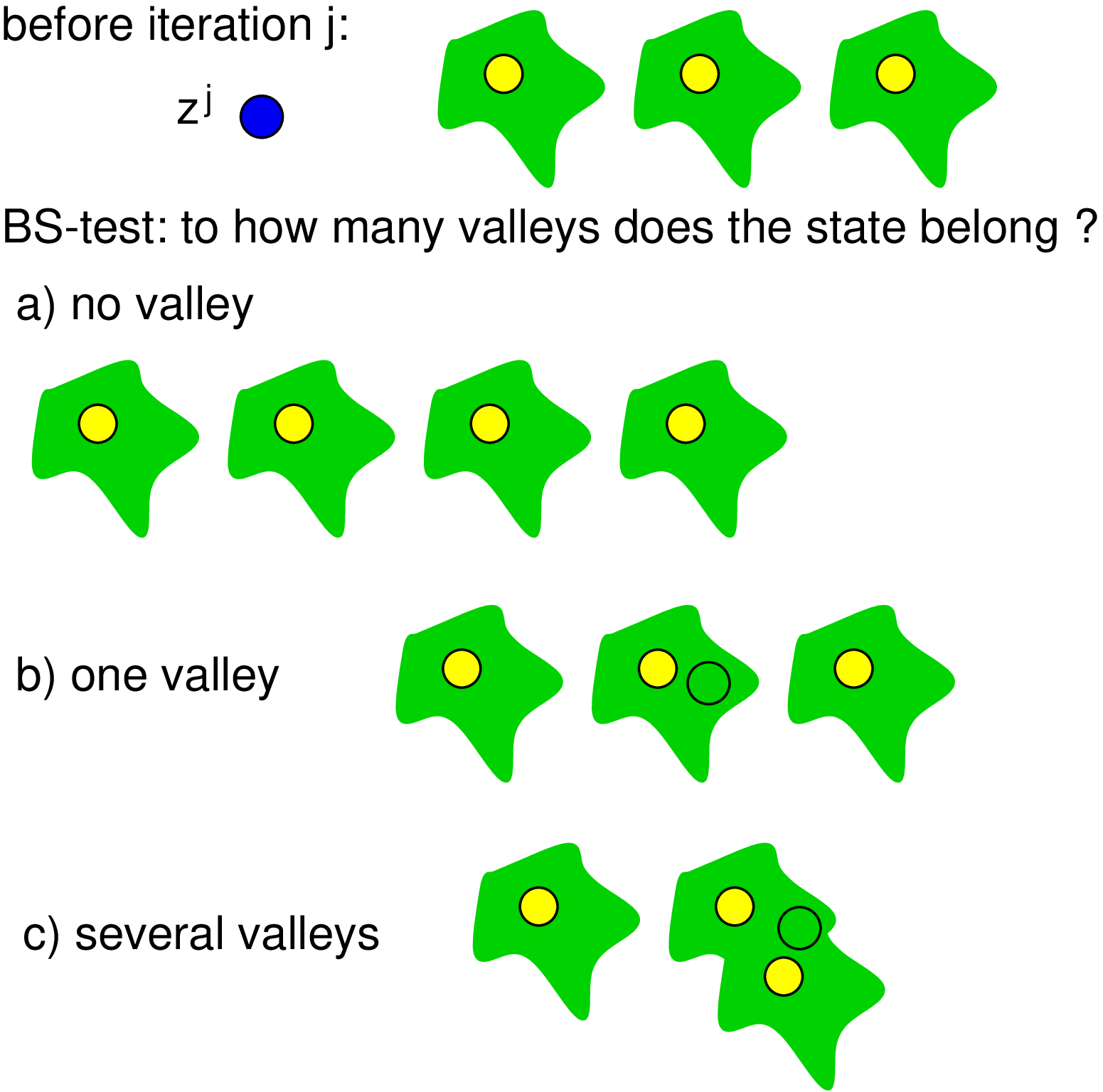}}
\end{center}
\caption{\captionThreeCases}
\label{figThreeCases}
\end{figure}

\begin{figure}[htb]
\begin{center}
\myscaleboxb{\includegraphics{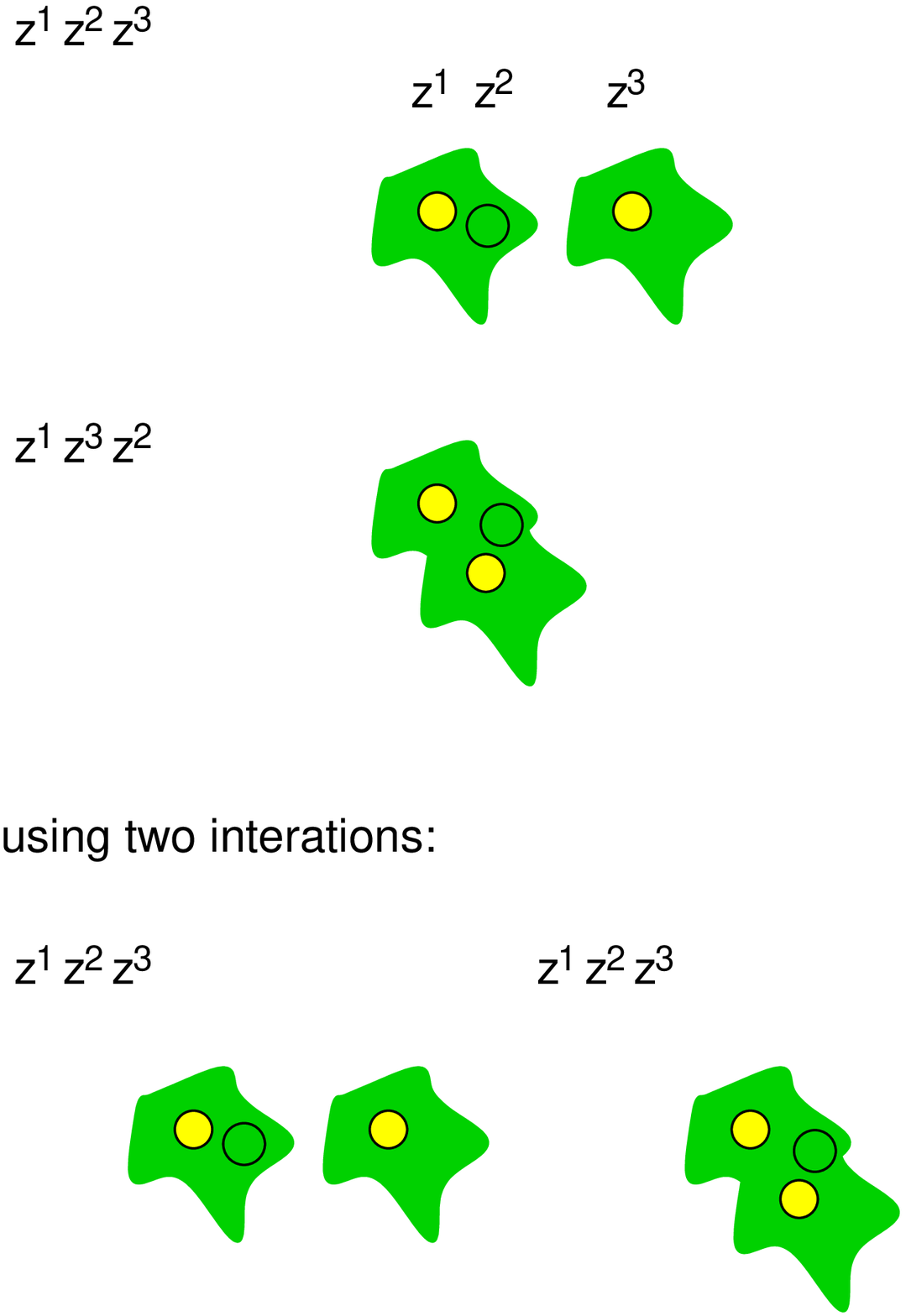}}
\end{center}
\caption{\captionTwoRuns}
\label{figTwoRuns}
\end{figure}

\begin{figure}[htb]
\begin{center}
\myscalebox{\includegraphics{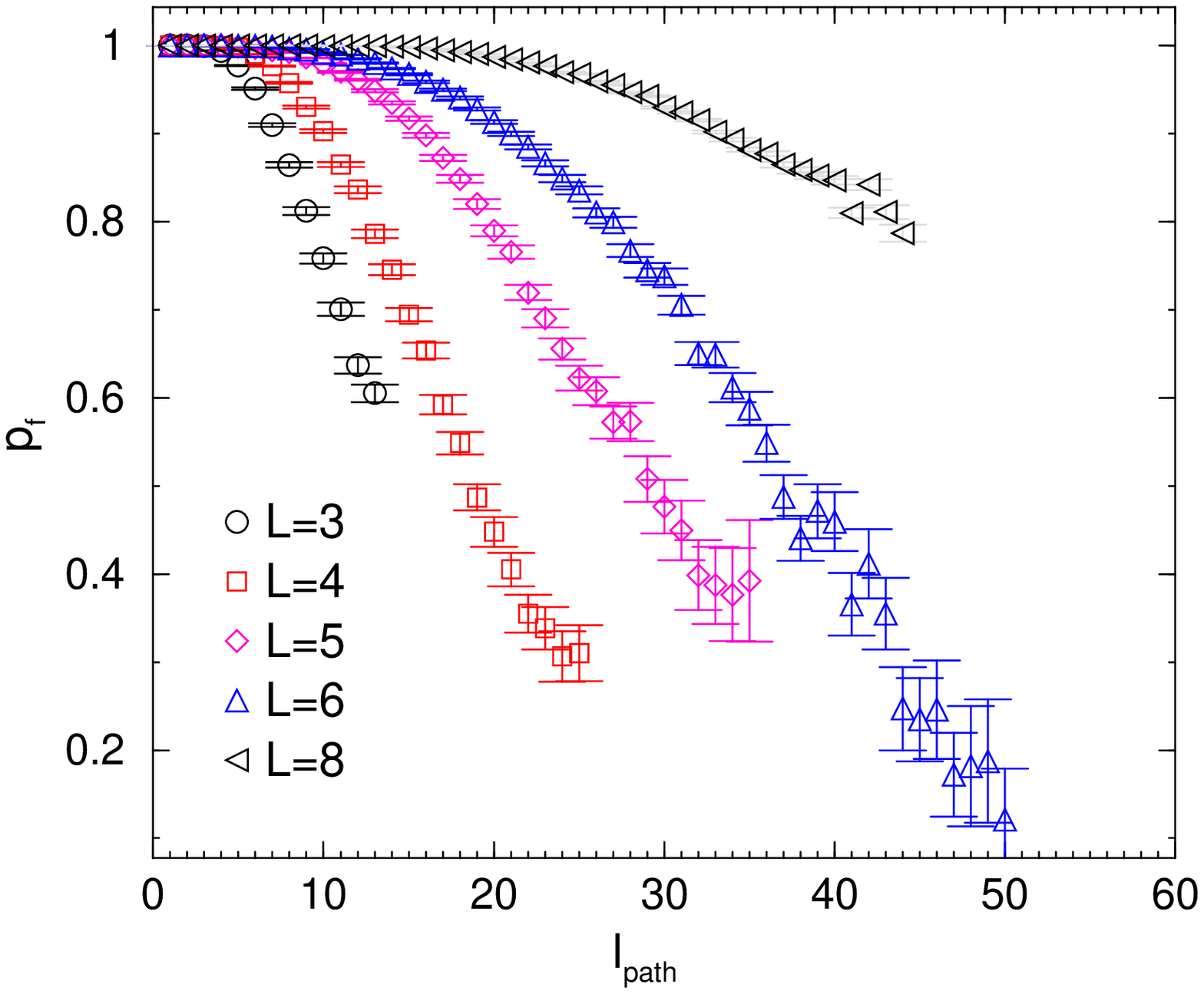}}
\end{center}
\caption{\captionPWayLength}
\label{figPWayLength}
\end{figure}

\begin{figure}[htb]
\begin{center}
\myscalebox{\includegraphics{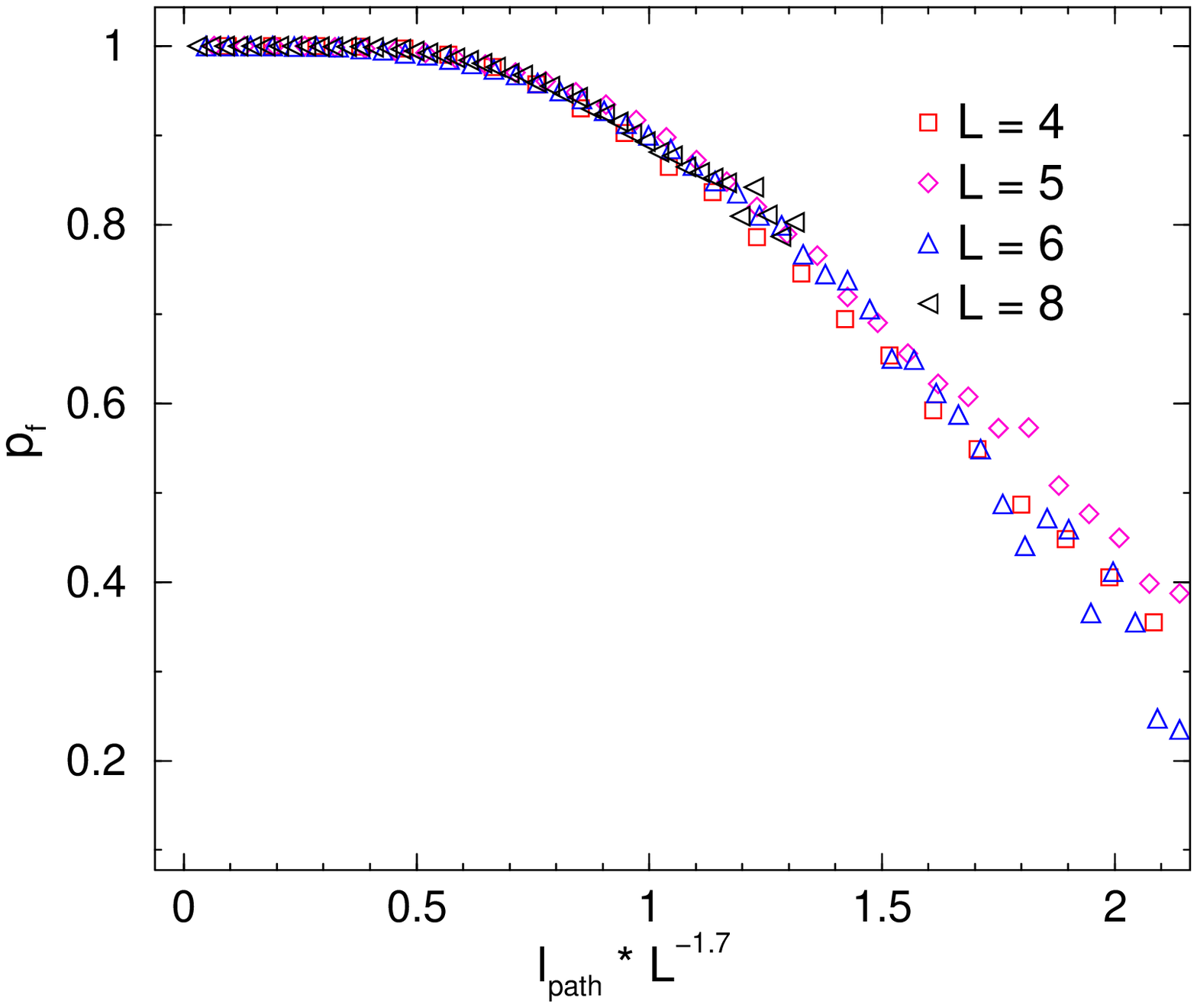}}
\end{center}
\caption{\captionPWayScale}
\label{figPWayScale}
\end{figure}

\begin{figure}[htb]
\begin{center}
\myscaleboxb{\includegraphics{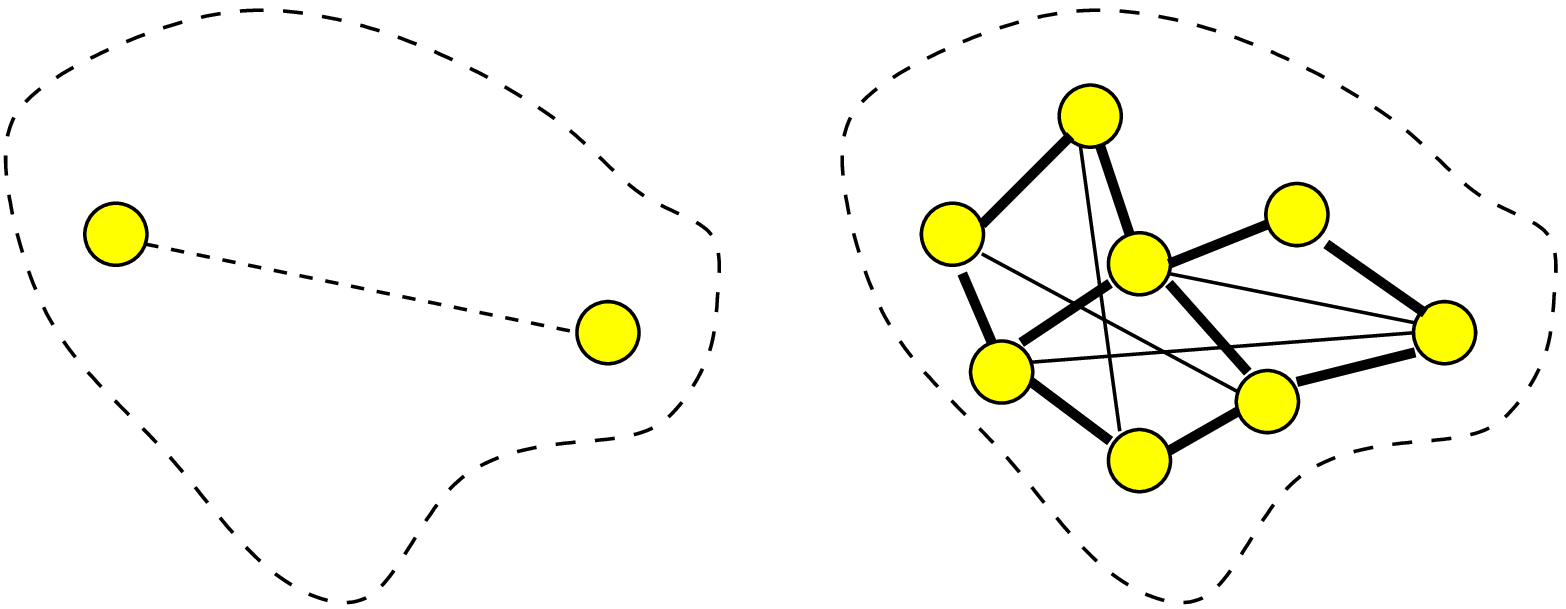}}
\end{center}
\caption{\captionMoreStates}
\label{figMoreStates}
\end{figure}

\begin{figure}[htb]
\begin{center}
\myscalebox{\includegraphics{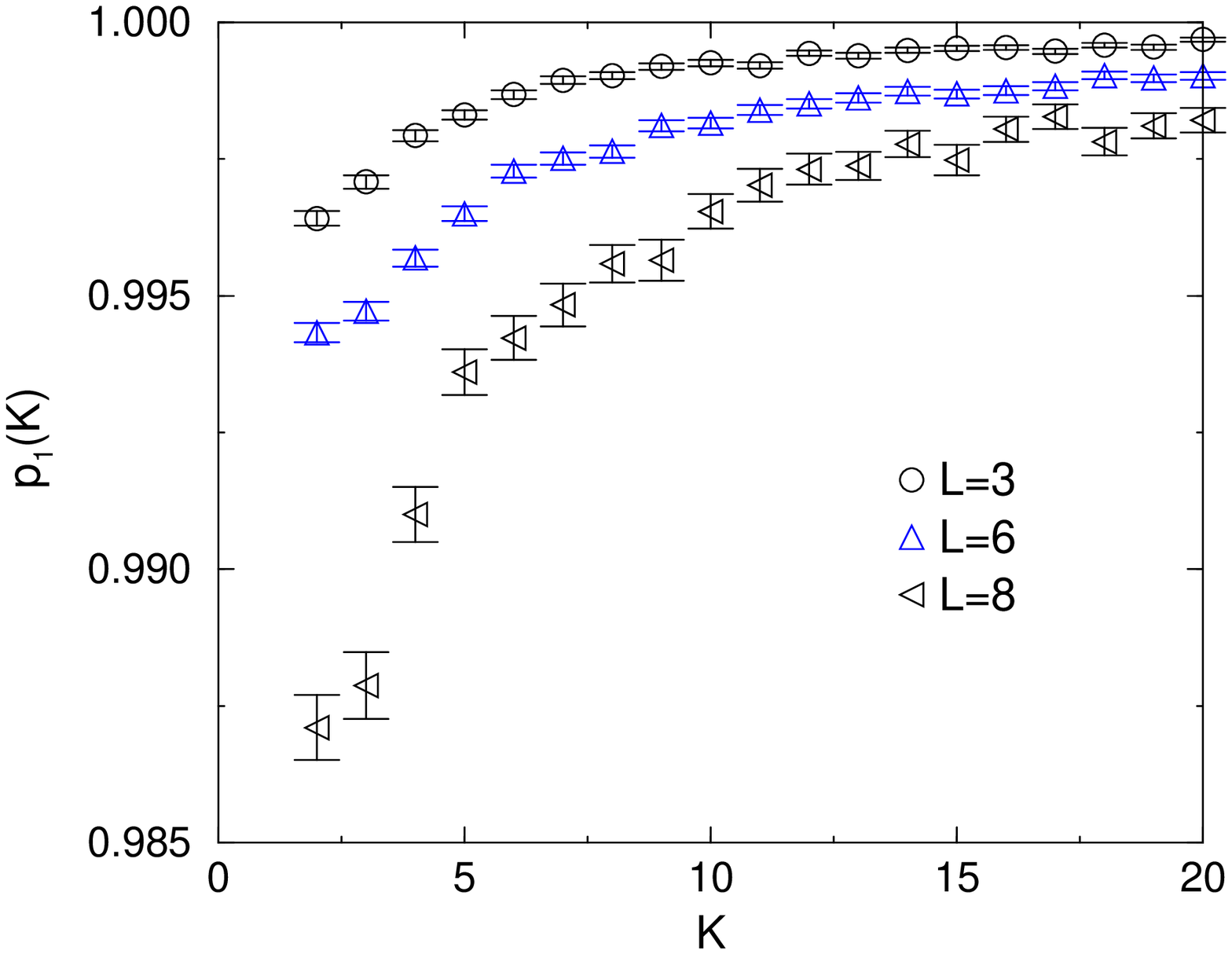}}
\end{center}
\caption{\captionPOneNum}
\label{figPOneNum}
\end{figure}

\begin{figure}[htb]
\begin{center}
\myscalebox{\includegraphics{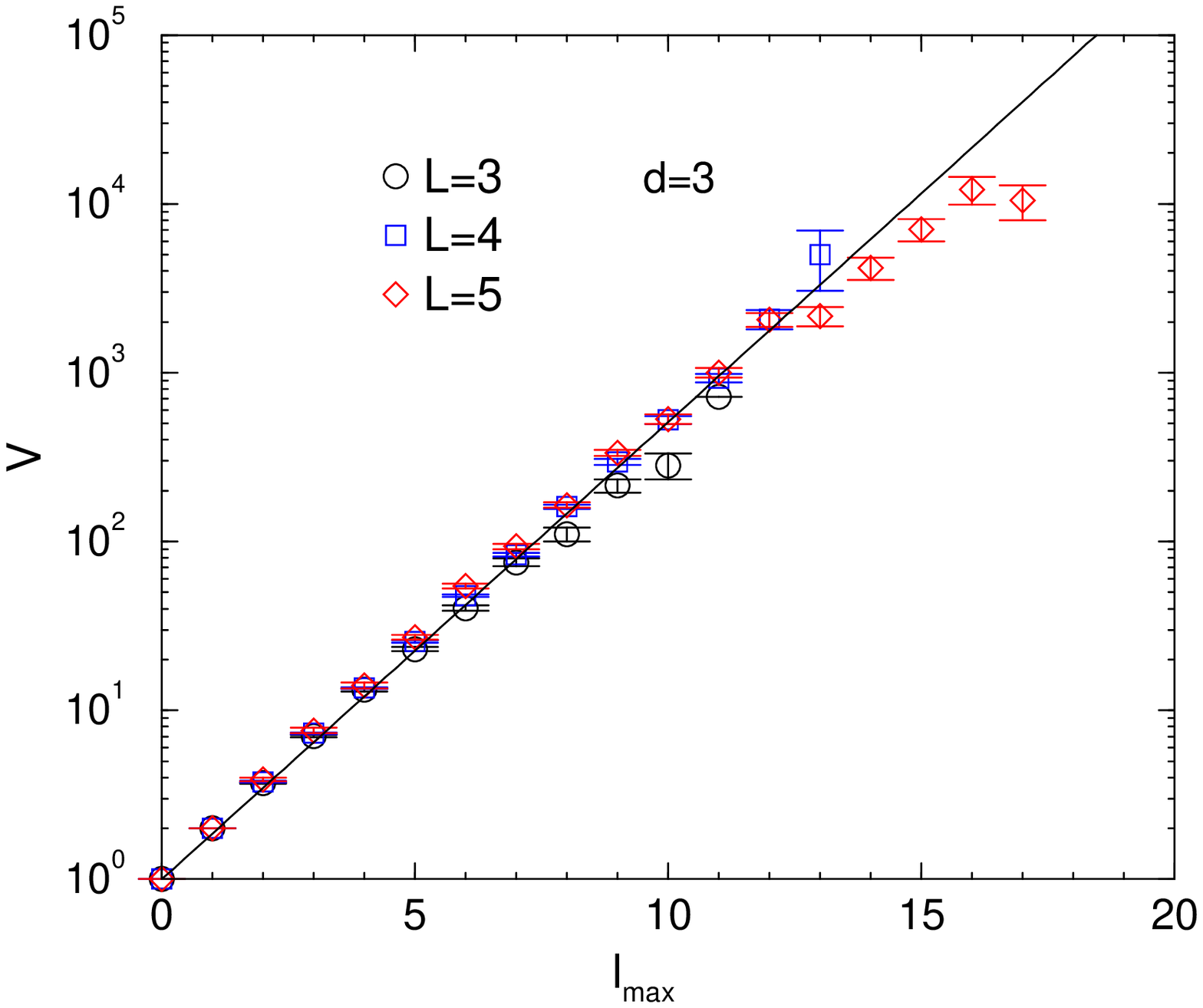}}
\end{center}
\caption{\captionEstSize}
\label{figEstSize}
\end{figure}

\begin{figure}[htb]
\begin{center}
\myscalebox{\includegraphics{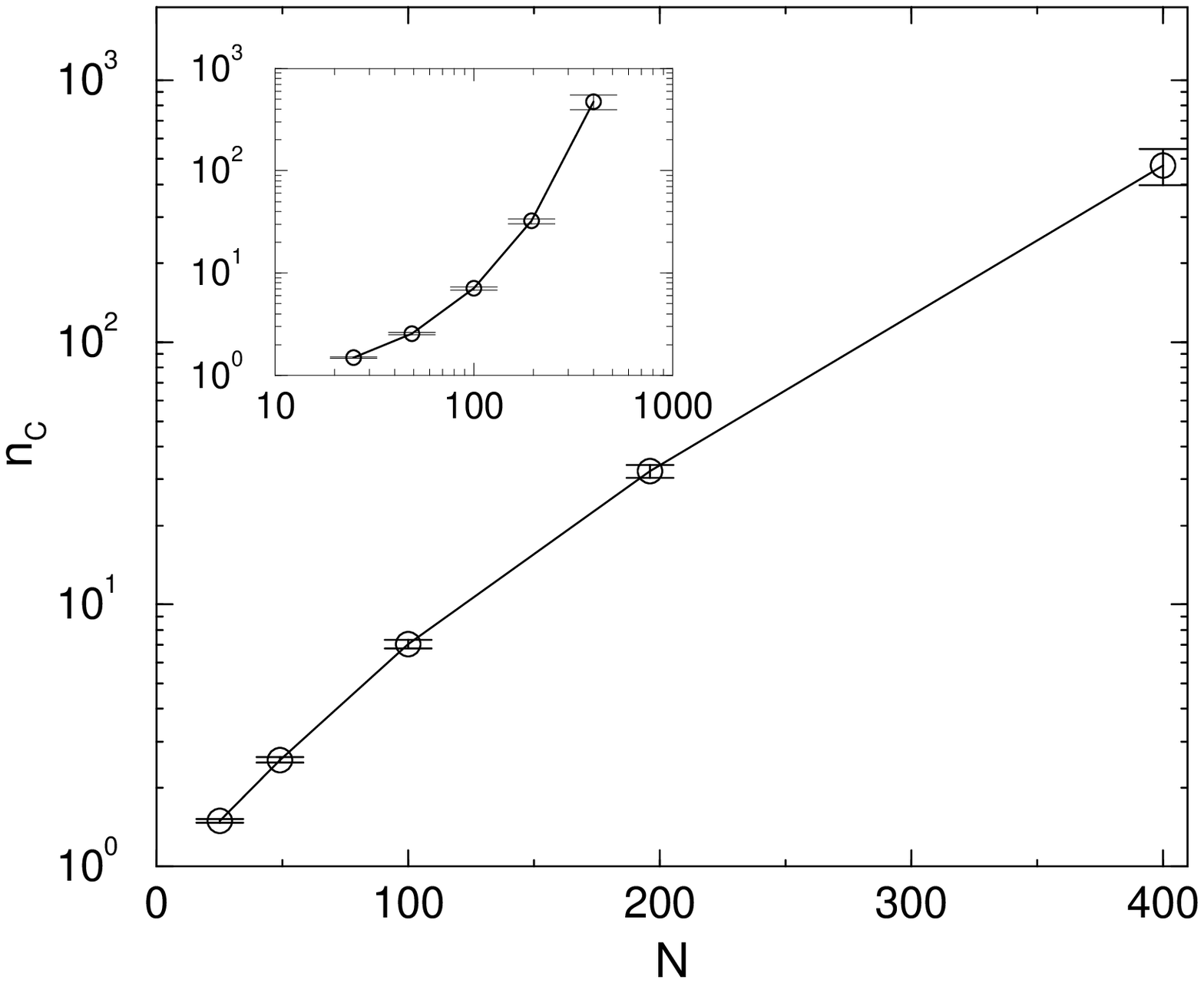}}
\end{center}
\caption{\captionNumClusters}
\label{figNumClusters}
\end{figure}

\begin{figure}[htb]
\begin{center}
\myscalebox{\includegraphics{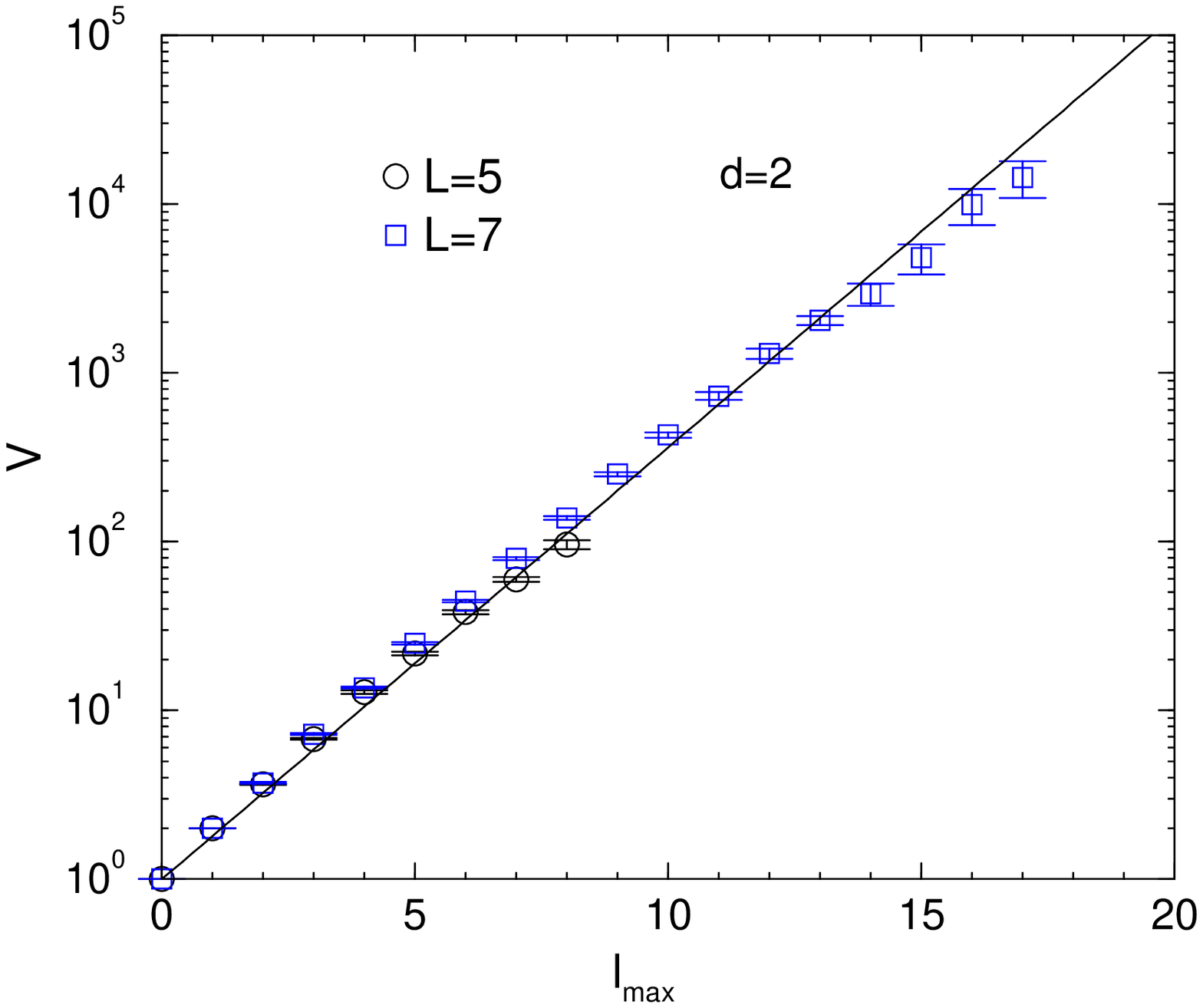}}
\end{center}
\caption{\captionEstSizeSquare}
\label{figEstSizeSquare}
\end{figure}

\begin{figure}[htb]
\begin{center}
\myscalebox{\includegraphics{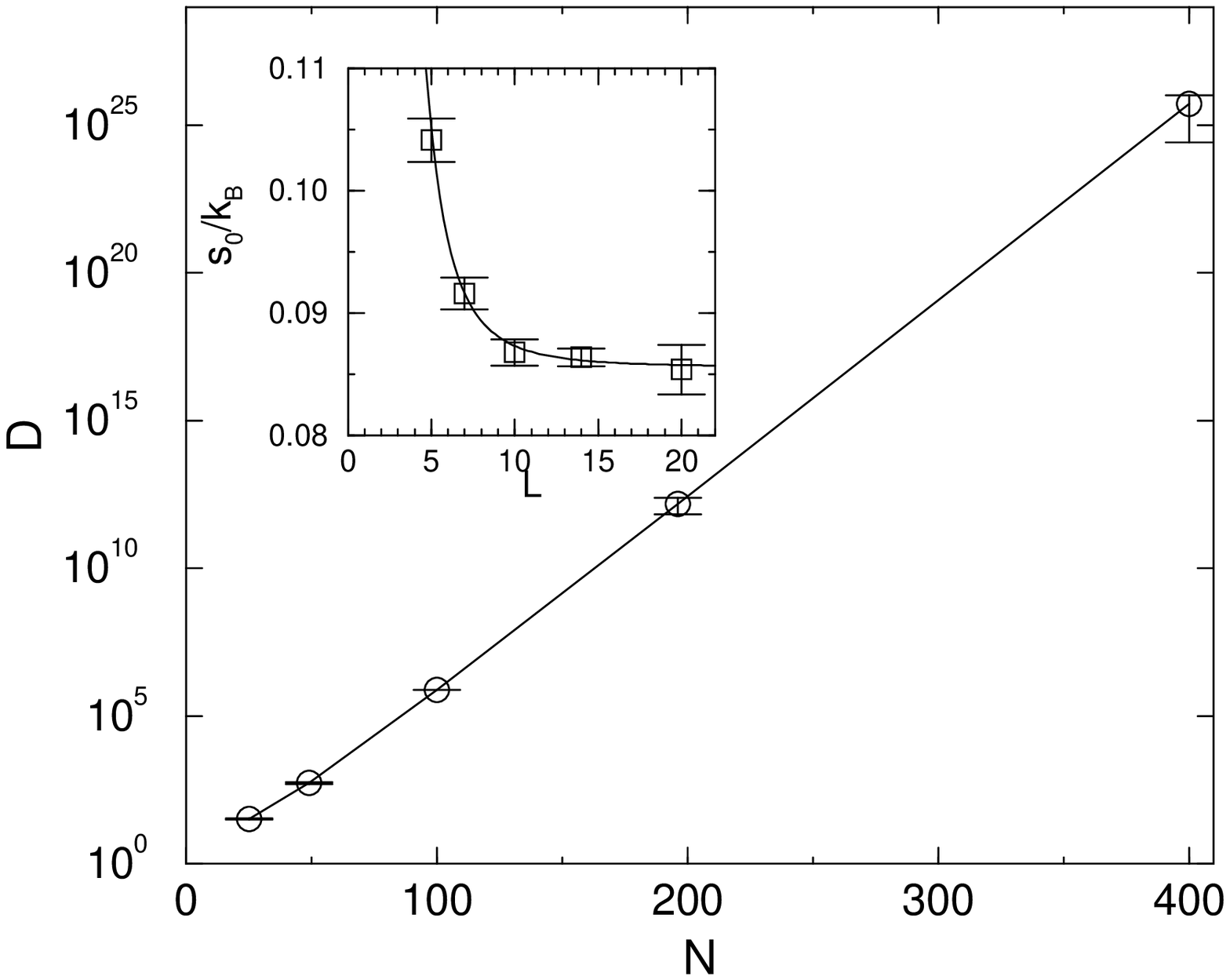}}
\end{center}
\caption{\captionNumStates}
\label{figNumStates}
\end{figure}

\begin{table}
\begin{tabular}{lrr}
L & $n_R$ & $r$ \\\hline
3 & 1000 & 1000 \\
4 & 1000 & $10^4$ \\
5 & 100 & $10^5$ \\
5 & 1000 & 3000 \\
6 & 1000 & 5000 \\
8 & 192 & $2\times 10^4$  
\end{tabular}
\caption{Three-dimensional $\pm J$ spin glasses. 
For each system size L: number $n_R$ of realizations, 
number $r$ of independent runs per realization.}
\end{table}


\begin{thebibliography}{99}

\bibitem{binder86} For reviews on spin glasses see: 
K. Binder and A.P. Young, Rev. Mod. Phys. {\bf 58}, 801 (1986);
 M. Mezard, G. Parisi, M.A. Virasoro, Spin
glass theory and beyond, World Scientific, Singapur 1987;
 K.H. Fisher and J.A. Hertz, Spin Glasses, Cambridge
University Press, 1991

\bibitem{alex-valleys} A.K. Hartmann, submitted (1999), cond-mat/9902120

\bibitem{rieger98} H. Rieger, in: Advances in Computer Simulation, 
     ed. J. Kertesz and I. Kondor,
    Lecture Notes in Physics {\bf 501}, Springer, Heidelberg, 1998

\bibitem{barahona82b} F. Barahona, R. Maynard, R. Rammal and
  J.P. Uhry, J. Phys. A {\bf 15}, 673 (1982).

\bibitem{barahona82} F. Barahona, J. Phys. A {\bf 15}, 3241 (1982)

\bibitem{hartwig84} A. Hartwig, F. Daske and S. Kobe, 
Comp. Phys. Commun. {\bf 32} 133 (1984)

\bibitem{klotz94} T. Klotz and S. Kobe, J. Phys. A {\bf 27}, L95  (1994)

\bibitem{simone95} C. De Simone, M. Diehl, M. J\"unger, P. Mutzel, 
G. Reinelt and G. Rinaldi, J. Stat. Phys. {\bf 80}, 487 (1995)

\bibitem{simone96} C. De Simone, M. Diehl, M. J\"unger, P. Mutzel, 
G. Reinelt and G. Rinaldi, J. Stat. Phys. {\bf 84}, 1363 (1996)

\bibitem{alex-stiff} A.K. Hartmann, Phys. Rev. E {\bf 59}, 84 (1999)

\bibitem{pal96} K.F. P\'al, Physica A {\bf 223}, 283 (1996)

\bibitem{michal92} Z. Michalewicz, Genetic Algorithms + Data Structures
= Evolution Programs, Springer, Berlin 1992

\bibitem{alex2} A.K. Hartmann, Physica A {\bf 224}, 480 (1996)



\bibitem{parisi2} G. Parisi, Phys. Rev. Lett. {\bf 43}, 1754 (1979); 
                  J. Phys. A {\bf 13}, 1101 (1980); {\bf 13}, 1887
                  (1980); {\bf 13}, L115 (1980);
                  Phys. Rev. Lett. {\bf 50}, 1946 (1983)


\bibitem{mcmillan} W.L. McMillan, J. Phys. C {\bf 17}, 3179 (1984)

\bibitem{bray} A.J. Bray and M.A. Moore, J. Phys. C {\bf 18}, L699 (1985)

\bibitem{fisher1} D.S. Fisher and D.A. Huse, Phys. Rev. Lett
         {\bf 56}, 1601 (1986).

\bibitem{fisher2} D.S. Fisher and D.A. Huse, Phys. Rev. B 
         {\bf 38} , 386 (1988).

\bibitem{bovier} A. Bovier and J. Fr\"ohlich, J. Stat. Phys. 
                 {\bf 44}, 347 (1986)

\bibitem{garstecki99} P. Garstecki, T.H. Hoang and M. Cieplak,
  preprint cond-mat/9904060

\bibitem{toulouse77} G. Toulouse, Commun. Phys. {\bf 2}, 115 (1977)

\bibitem{claibo} J.D. Claiborne, Mathematical Preliminaries for
Computer Networking, Wiley, New York 1990

\bibitem{knoedel} W. Kn\"odel, Graphentheoretische Methoden und ihre
Anwendung Springer, Berlin 1969

\bibitem{swamy} M.N.S. Swamy and K. Thulasiraman, Graphs, Networks and
Algorithms (Wiley, New York 1991)

\bibitem{picard1} J.-C. Picard and H.D. Ratliff, Networks {\bf 5}, 357 (1975)

\bibitem{traeff} J.L. Tr\"aff, Eur.\ J.\ Oper.\ Res.\ {\bf 89}, 564 (1996)

\bibitem{tarjan} R.E. Tarjan, Data Structures and Network Algorithms,
Society for industrial and applied mathematics, Philadelphia 1983

\bibitem{alex-2d} A.K. Hartmann, Eur. Phys. J. B {\bf 8}, 619 (1999)

\bibitem{alex-sg2} A.K. Hartmann, Europhys. Lett. {\bf 40}, 429 (1997); 
Europhys. Lett. {\bf 44}, 249 (1998); 
Europhys. Lett. {\bf 45}, 747 (1999) 

\bibitem{alex-threshold} A.K. Hartmann, Phys. Rev. B {\bf 59}, 3617 (1999)

\bibitem{hoshen76} J. Hoshen and R. Kopelman, Phys. Rev. B {\bf 14},
  3438 (1976)

\bibitem{alex-eig-cea} A.K. Hartmann, submitted, preprint 
cond-mat/9904364 (1999) 

\bibitem{morgenstern80b} I. Morgenstern and K. Binder, Phys. Rev. B, {\bf
    22}, 288 (1980)

\bibitem{vannimenus77} J. Vannimenus and G. Toulouse, J. Phys. C {\bf 10},
    L537 (1977)

\bibitem{kirkpatrick77} S. Kirkpatrick, Phys. Rev. B {\bf 16}, 4630 (1977)


\end{thebibliography}
\end{document}